# Fine structure of soliton bound states in the parametrically driven, damped nonlinear Schrödinger equation

M.M. Bogdan[1], O.V. Charkina[1,2]

[1]*B. Verkin Institute for Low Temperature Physics and Engineering of the National Academy of Sciences of Ukraine, 47 Nauky Ave., Kharkiv, 61103, Ukraine*

[2]*University of Luxembourg, L-1511 Luxembourg City, Luxembourg*

E-mail: charkina@ilt.kharkov.ua
bogdan@ilt.kharkov.ua

**Abstract**

Static soliton bound states in nonlinear systems are investigated analytically and numerically in the framework of the parametrically driven, damped nonlinear Schrödinger equation. We find that the ordinary differential equations, which determine bound soliton solutions, can be transformed into the form resembling the Schrödinger-like equations for eigenfunctions with the fixed eigenvalues. We assume that a nonlinear part of the equations is close to the reflectionless potential well occurring in the scattering problem, associated with the integrable equations. We show that symmetric two-hump soliton solution is quite well described analytically by the three-soliton formula with the fixed soliton parameters, depending on the strength of parametric pumping and the dissipation constant.

## Introduction

Progress of the theory of nonlinear phenomena is directly related to the emergence of the soliton concept. Academician A.S. Davydov is one of the founders of the nonlinear science schools in Ukraine. The Davydov soliton is a bright example of the application of the soliton concept to the nonlinear transport phenomena of the quantum quasiparticles in biophysics [1, 2]. The theoreticians of Davydov's scientific school continues the best traditions of their teacher, describing a wide range of the soliton effects in biophysics and condensed matter [3, 4]. Exploiting the soliton description, based on the use of the nonlinear Schrödinger equation and its modifications, Davidov's school has created a new direction of investigation of nonlinear phenomena. Remembering with deep respect A.S. Davydov as a true science classic, we dedicate this paper to his memory on the 110th anniversary of his birth.

One of the generalizations of the nonlinear Schrödinger equation arises when the parametric pumping and dissipation are taken into account. At first, the corresponding three-dimensional equation appeared in the theory of parametric instability of spin waves [5]. The one-dimensional variant of the equation has been used for description of dynamics of the parametrically driven soliton (breather) in low-dimensional ferromagnets [6, 7]. The 1D equation is also applicable to a large variety of weakly nonlinear physical systems, beginning from the long channel with water under conditions of the Faraday resonance, where a non-propagating hydrodynamic soliton has been observed [8, 9], and up to nonlinear fibers with optical solitons [10]. The equation has been called the parametrically driven, damped nonlinear Schrödinger equation (PDNLSE) [7]. This equation has become a useful tool for the numerical study of regular and chaotic internal dynamics of the parametrically driven soliton under the dissipation [11]. In the framework of the PDNLSE the oscillatory instability of the soliton has been revealed and explained. The soliton stability diagram has been constructed, indicating a rich pattern of routes to chaos in the soliton internal dynamics in accordance with Feigenbaum and quasi-periodic scenarios [11]. However, the variational approach to the description of the internal soliton dynamics could not give a quantitative coincidence with numerical results for the soliton stability boundary [see, e.g., 12]. It means the internal structure of the stable parametric soliton may be more complicated than it seems. It has been shown [13] that two identical well-separated parametric solitons in the case of absence of the dissipation attract each other. Hence, they rather transform into a single stable soliton under influence of the dissipation than they could form a static soliton bound state [14]. At the same time in Ref. 15 the attempt to explore the interactions of solitons has led to the conclusion on the principal absence of bound states of two identical solitons in the PDNSLE. The fact, that this is wrong, became apparent soon after their evidence in the hydrodynamic Faraday resonance experiments [16, 17] and later, when the stable bound states have been found directly in numerical simulations of the PDNLSE [18]. Much earlier double soliton structures with well-separated humps have been observed experimentally in binary fluid mixtures [19, 20].

In the present work, we study the fine structure of the symmetric double-hump solutions of the PDNLSE. At first, we formulate this equation in terms of magnetism and reduce it to an autonomic one, namely called the PDNLSE, and adduce its exact stable and unstable one-soliton solutions and review numerical and analytical results concerning the stable double-hump soliton solutions. Then we propose an analytical approach to solving the governing system of ordinary differential equations in order to reproduce the static solutions found numerically. We show that the double-hump solution is described quite well by the three-soliton formula, typical for the integrable equations, with the fixed parameters, depending on the strength of parametric pumping and the dissipation constant. In conclusion, we discuss a possibility of a new ansatz in form of the multisoliton solution in order to describe stability properties and complex internal dynamics of the soliton bound states in the PDNLSE.

# Exact static soliton solutions of the PDNLSE and its reductions

The Landau-Lifshitz equation for the magnetization, describing the quasi-one-dimensional easy-plane anisotropic ferromagnet with a magnetic field in the easy plane, can be reduced in the weakly nonlinear limit to the dimensionless equation, as shown in Refs. 6 and 7:

$$i\Psi_t + \Psi_{xx} + 2|\Psi|^2 \Psi = h\exp(2it)\Psi^* - i\gamma\Psi \quad . \tag{1}$$

Here the real and imaginary parts of the function $\Psi(x,t)$ are proportional to small deviations of the unit magnetization vector $\vec{\mathbf{m}}(x,t)$, namely projections $m_z$ and $m_y$, respectively, from the direction of the constant magnetic field, lying in the easy-plane and being parallel to the alternating field. In [6, 7] all the relations are presented between the dimensionless time $t$, coordinate $x$ along the chain, the strength of parametric pumping $h$ and the dissipation coefficient $\gamma$ and corresponding dimensional parameters, including the anisotropy constant, the constant magnetic field, the alternating microwave field and the relaxation constant. Asterisk denotes the complex conjugate variable.

The parametric character of the driving force is evident because the resonant oscillation in the system occurs at half of the pumping frequency. After substitution $\Psi(x,t) = \psi\exp(it)$ to Eq. (1), we obtain the autonomic equation

$$i\psi_t + \psi_{xx} - \psi + 2|\psi|^2 \psi = h\psi^* - i\gamma\psi \quad , \tag{2}$$

which is called the PDNLSE and describes many nonlinear phenomena under parametric resonance conditions. The resonant nonlinear oscillation behaves as a small amplitude breather with a steady profile $\psi(x)$, which is exponentially localized in space, as it is clearly seen, e.g., on a wall of a water tank in the hydrodynamic experiments under the Faraday resonance [9]. The exact one-soliton solutions of Eq. (2) are well-known [5 – 8]:

$$\psi_+(x) = \frac{\kappa_+}{\cosh(\kappa_+ x)}\exp(-i\theta), \tag{3}$$

$$\psi_-(x) = -i\frac{\kappa_-}{\cosh(\kappa_- x)}\exp(i\theta), \tag{4}$$

where the parameters of soliton amplitudes and phase are fixed as

$$\kappa_\pm = \sqrt{1 \pm \sqrt{h^2 - \gamma^2}}, \quad \theta = \frac{1}{2}\arcsin(\frac{\gamma}{h}), \tag{5}$$

It has been shown in Ref. 7 that soliton $\psi_-(x)$ is absolutely unstable, and soliton $\psi_+(x)$ is stable in the definite region on the plane of the parameters $h$ and $\gamma$. Outside the stability domain it becomes unstable with increasing the driving force, because of the resonance of its two internal modes [7]. Such a scenario of the emergence of the instability mode in the internal soliton dynamics in continuous and discrete systems is now known as oscillatory instability. Above the found stability boundary, coming through the cascade of bifurcations, the soliton $\psi_+(x)$ demonstrates a transition to temporary chaos in its internal dynamics, preserving spatial coherence [11]. Besides this Feigenbaum scenario, the soliton can also undergo destruction through the quasi-periodic route to chaos.

In the dissipation-free case, the system of ordinary differential equations, determining any static solution $\psi(x) = \psi_1 + i\psi_2$, takes the form

$$\left(\frac{d^2}{dx^2} + 2(\psi_1^2 + \psi_2^2)\right)\psi_1 = \mu_+^2 \psi_1, \qquad \left(\frac{d^2}{dx^2} + 2(\psi_1^2 + \psi_2^2)\right)\psi_2 = \mu_-^2 \psi_2, \qquad (6)$$

where parameters $\mu_\pm = \sqrt{1 \pm h}$. When $\gamma = 0$ and, hence, the parameter $\theta = 0$, the solutions (3) and (4) become purely real and imaginary functions, respectively. In this case, they are trivial one-component soliton solutions because the second component is equal to zero identically. There is the non-trivial solution of the system (6) with components in the form of bright and dark solitons:

$$\psi_1 = \sqrt{\mu_+^2 - \frac{1}{2}\mu_-^2}\, \text{sech}(qx), \qquad \psi_2 = \sqrt{\frac{1}{2}(\mu_-^2 - q^2)}\, \tanh(qx), \qquad (7)$$

where $q = \sqrt{\mu_+^2 - \mu_-^2} = \sqrt{2h}$. In result, the corresponding doubled density $\rho = 2(\psi_1^2 + \psi_2^2)$ can be written in the form $\rho = \mu_-^2 + 2q^2 \text{sech}^2(qx)$, which makes the functions (7) obvious solutions of the system (6).

For the case of the fixed value of the parameter $h = h_0 = 3/5$, i.e., when $\mu_- = \mu_+/2 = \sqrt{2/5}$, there is the two-component solution of the form [21]

$$\psi_1 = \sqrt{3}\mu_- \text{sech}^2(\mu_- x), \qquad \psi_2 = \sqrt{3}\mu_- \tanh(\mu_- x)\text{sech}(\mu_- x). \qquad (8)$$

Moreover, it has been found out that the system of coupled equations (6) is completely integrable because it possesses two independent integrals of motion [22]. In Ref. 4, using the elliptic coordinates presentation of the components $\psi_1$ and $\psi_2$, the solution of the system has been written in quadratures. As a result, it has been shown, that in addition to the trivial solitons (3) and (4), and besides special ones (7) and (8), there is the general form of the double-soliton solutions, named the bisolitons [4], which complete the set of localized solutions. They correspond to the specific choice of values of integrals of motion of Eqs. (6), that ensures vanishing the soliton solutions at infinity.

For the first time, the explicit form of double solitons of the system (6) has been found in Ref. 21 by the Hirota direct method for finding the soliton solutions [23]. After the substitution to Eqs. (6) the function transformation:

$$\psi_1 = \frac{\varphi_1}{\Phi}, \qquad \psi_2 = \frac{\varphi_2}{\Phi}, \qquad (9)$$

the system can be reduced to the Hirota bilinear equations

$$D_x^2 \varphi_1 \cdot \Phi = \mu_+^2 \varphi_1 \cdot \Phi, \quad D_x^2 \varphi_2 \cdot \Phi = \mu_-^2 \varphi_2 \cdot \Phi, \quad D_x^2 \Phi \cdot \Phi = 2(\varphi_1^2 + \varphi_2^2), \qquad (10)$$

where the new functions $\varphi_1$ and $\varphi_2$ are assumed to relate to the denominator function $\Phi$ by the third equation and the Hirota differential operator is defined as follows

$$D_x^2 \varphi_1 \cdot \varphi_2 = \left( \frac{\partial}{\partial x} - \frac{\partial}{\partial x'} \right)^2 \varphi_1(x)\varphi_2(x') \bigg|_{x=x'}. \qquad (11)$$

The soliton solutions of Eqs. (6) are expressed in terms of the standard finite series of exponents [23]. They can be written as symmetric and antisymmetric functions in the explicit form:

$$\psi_1 = \frac{2\mu_+ \sqrt{\sigma} \cosh(\mu_- x)}{\cosh((\mu_+ + \mu_-)x) + \sigma \cosh((\mu_+ - \mu_-)x)}, \qquad (12)$$

$$\psi_2 = \frac{2\mu_- \sqrt{\sigma} \sinh(\mu_+ x)}{\cosh((\mu_+ + \mu_-)x) + \sigma \cosh((\mu_+ - \mu_-)x)}, \qquad (13)$$

where the parameter $\sigma = (\mu_+ + \mu_-)/(\mu_+ - \mu_-)$. As noted in Ref. 21, every equation in the system (6) can be considered as the Schrödinger-like equation of the corresponding eigenvalue problem with the symmetric "potential" $V \equiv -\rho = -2(\psi_1^2 + \psi_2^2)$. Continuing this analogy, we introduce the normalized eigenfunctions $\tilde{\psi}_1 = \psi_1 / \sqrt{2\mu_+}$ and $\tilde{\psi}_2 = \psi_2 / \sqrt{2\mu_-}$. Then the potential, written in terms of the eigenfunctions, takes the form $V = -4(\mu_+ \tilde{\psi}_1^2 + \mu_- \tilde{\psi}_2^2)$. The value of the potential at the origin is expressed through the eigenvalues as follows: $V(0) = -2(\mu_+^2 - \mu_-^2) = -4h$. It is easy to see that the potential well has two equivalent minima for the parameter values of $0 < h < 1/2$ and only one minimum for $1/2 \le h < 1$, and when $h = 1/2$ then $U(0) = -2$.

Finally note, that having the double-hump profile, the quantity $|\psi(x)|^2 = \psi_1^2 + \psi_2^2$ can be considered either as the two-peak localized density of spin excitations [6] or interpreted as the electric bisolitons in molecular chains [4], depending on physical applications.

**Three-soliton approach to bound states in damped systems with parametric pumping**

The numerical investigations of the PDNLSE, stimulated by the hydrodynamic Faraday resonance experiments [16 – 18], have revealed, besides the complex internal one-soliton dynamics, also the formation of stable complexes of two solitons and their mutual dynamical transformation [24 – 28]. The direct approach to finding the stable static double-hump solitons consists in simulation of the autonomic PDNLSE with suggesting that the parameters $h$ and $\gamma$ lie in the soliton stability region [26]. The stable solitons manifest themselves as stationary attractors in the dynamics simulation.

Here we concentrate only on the fact of the existence of the static solutions of the PDNLSE for symmetric double-hump solitons. Both stable and unstable bound soliton configurations can be found numerically by solving the following system of nonlinear ordinary equations

$$\left(\frac{d^2}{dx^2} + 2(\psi_1^2 + \psi_2^2)\right)\psi_1 = \mu_+^2 \psi_1 - \gamma \psi_2 , \tag{14}$$

$$\left(\frac{d^2}{dx^2} + 2(\psi_1^2 + \psi_2^2)\right)\psi_2 = \mu_-^2 \psi_2 + \gamma \psi_1 . \tag{15}$$

Unfortunately, the analytical description of these solutions by the use of the variational approach proposed in [27] appeared to be unsatisfactory quantitatively. Here we aim to find an appropriate analytical approximation for the static complex solutions. Before doing that, we note the main lack of the variational ansatz proposed in Ref. [27]. Starting with the fact that the lonely soliton $\psi_+(x)$ is stable and the lonely soliton $\psi_-(x)$ is absolutely unstable, the authors of [27] neglected the latter at all. At the same time, it is easy to see that the asymptotics of the found double-hump numerical solutions decay more slowly than those of $\psi_+(x)$ but exactly like those of $\psi_-(x)$. That is why in order to deal with the bound soliton solution, we consider a general structure of the static solutions

$$\psi(x) = \psi_1 + i\psi_2 = u \exp(-i\theta) - iv \exp(i\theta) \tag{16}$$

by the introduction of the new functions $u(x)$ and $v(x)$ instead $\psi_1$ and $\psi_2$, where the parameter $\theta$ is defined by Eq. (5). As seen from Eq. (16), we consider the generalization for any static solution as a simple sum or a superposition of two types of solitons, as it would be in the case of single solitons $\psi_+(x)$ and $\psi_-(x)$ (compare Eq. (16) with Eqs. (3) and (4)). However, in fact, we present the real and imaginary parts of the function $\psi(x)$ in the form of the linear combination of new functions $u$ and $v$ as follows:

$$\psi_1 = u\cos\theta + v\sin\theta , \qquad (17)$$

$$\psi_2 = -u\sin\theta - v\cos\theta , \qquad (18)$$

in order to exclude the coupling of the equations (14) and (15) by the linear terms. At first glance, the transformation is similar to the rotation, but the corresponding matrix is not unitary,

$$\mathbf{Y} = \begin{pmatrix} \cos\theta & \sin\theta \\ -\sin\theta & -\cos\theta \end{pmatrix} \qquad (19)$$

because its determinant $\det\mathbf{Y} = -\cos 2\theta$. However, it is easy to find that the inverse transformation of the above functions is the following:

$$u = (\psi_1 \cos\theta + \psi_2 \sin\theta)/\cos 2\theta , \qquad (20)$$

$$v = -(\psi_1 \sin\theta + \psi_2 \cos\theta)/\cos 2\theta , \qquad (21)$$

therefore, the inverse matrix $\mathbf{Y}^{-1} = \mathbf{Y}/\cos 2\theta$, i.e., the matrix $\mathbf{M} = \mathbf{M}^{-1} = \mathbf{Y}/\sqrt{\cos 2\theta}$ and $\mathbf{M}^2 = \mathbf{I}$, where $\mathbf{I}$ is the identity matrix. After the transformation (17) and (18), instead of Eqs. (14) and (15), we obtain the following system of the Schrödinger-like equations:

$$\left(\frac{d^2}{dx^2} + 2(u^2 + \lambda uv + v^2)\right)u = \kappa_+^2 u , \qquad (22)$$

$$\left(\frac{d^2}{dx^2} + 2(u^2 + \lambda uv + v^2)\right)v = \kappa_-^2 v , \qquad (23)$$

where the parameter $\lambda = 2\gamma/h$. Indeed, if we find the solutions $u$ and $v$ and put them in the nonlinear part, and introduce the effective potential

$$U = -2(u^2 + \lambda uv + v^2) , \qquad (24)$$

then we can present Eqs. (22) and (23) as the equations of the spectral problem for the Schrödinger operator

$$L = -\frac{d^2}{dx^2} + U , \qquad (25)$$

$$Lu = -\kappa_+^2 u, \qquad Lv = -\kappa_-^2 v . \qquad (26)$$

Thus it appears that the function $u(x)$ is the operator eigenfunction, corresponding to the eigenvalue $-\kappa_+^2$, and the function $v(x)$ is the eigenfunction, corresponding to the eigenvalue $-\kappa_-^2$, respectively. Note that the explicit expressions of the eigenvalues are given by Eq. (5). Now it is clear that many arguments, which have been relevant in the case of the system (6), can be used to analyze the properties of the solutions $u(x)$ and $v(x)$.

However, from the very beginning, it is hard to expect that the initial system of Eqs. (14) and (15), and hence, the system (26), could be solved exactly because now no integrals of motion are available, which allowed finding the complete solution of the dissipation-free problem. Nevertheless, it is possible to imagine what kind of localized solutions the system (26) possesses. Since we are interested in the symmetric potential $U(x)$ and $\kappa_- < \kappa_+$, then the solution $u(x)$ can be the even eigenfunction of a ground state of the Schrödinger operator $L$, having no zeroes. The function $v(x)$ can also appear to be the even eigenfunction of the operator, but then it has to correspond to the second excited state of the operator $L$. Therefore, in contrast to Eqs. (6), the potential well (24) can contain three discrete levels and, hence, three discrete eigenvalues can exist. The eigenvalue $-\kappa^2$, corresponding to the odd eigenfunction $\chi(x)$ and to the first excited state, is placed between two eigenvalues of Eqs. (26) and hence $\kappa_- < \kappa < \kappa_+$. Surely, because of the nonlinearity of the problem (22) – (23), the amplitudes of the functions $u(x)$ and $v(x)$ have definite values, whereas the function $\chi(x)$ is considered simply as the eigenfunction of the operator $L$ with the effective potential $U(x)$ and it can be normalized to unity:

$$L\chi = -\kappa^2 \chi , \qquad \int_{-\infty}^{\infty} \chi^2 dx = 1 . \qquad (27)$$

The potential well $U(x)$ can be always presented as follows

$$U(x) = -2\frac{d^2}{dx^2}\ln F = -\frac{D_x^2 F \cdot F}{F^2}, \qquad (28)$$

where the function $F(x)$ is found by the consequent integration of the equation

$$\frac{d}{dx}\left(\frac{1}{F}\frac{dF}{dx}\right) = u^2 + \lambda uv + v^2 , \qquad (29)$$

if the functions in the right-hand side of Eq. (29) are known. Then we introduce the functions $g$ and $f$ as follows

$$g = uF, \qquad f = vF , \qquad (30)$$

and transform Eq. (29) into the Hirota bilinear equation

$$D_x^2 F \cdot F = 2(g^2 + \lambda g f + f^2) , \tag{31}$$

and finally, reduce Eqs. (22) and (23) to the standard form of the bilinear equations:

$$D_x^2 g \cdot F = \kappa_+^2 g \cdot F, \qquad D_x^2 f \cdot F = \kappa_-^2 f \cdot F . \tag{32}$$

Comparing the system of Eqs. (31) and (32) and the system (10), we see that the principal difference between them consists in the presence of the term with dissipation coefficient in the r.h.s. of Eq. (31). This does not allow to solve exactly Eqs. (31) and (32) by the Hirota direct method and to find the solutions in the standard form of finite series of exponents [23].

Nevertheless, we can test the proximity of the numerical solutions of Eqs. (22) and (23) to exact solutions of the Hirota bilinear equations, which are true for the case of three solitons [23, 29]. We write the three-soliton bilinear equations in the form:

$$D_x^2 \varphi_+ \cdot F_0 = \kappa_+^2 \varphi_+ \cdot F_0, \quad D_x^2 \varphi \cdot F_0 = \kappa^2 \varphi \cdot F_0, \quad D_x^2 \varphi_- \cdot F_0 = \kappa_-^2 \varphi_- \cdot F_0 , \tag{33}$$

$$D_x^2 F_0 \cdot F_0 = 2(\varphi_+^2 + \varphi^2 + \varphi_-^2), \tag{34}$$

where the explicit forms of their functions $\varphi_+$, $\varphi$, and $\varphi_-$ are as follows

$$\varphi_+ = A_+ \left( \cosh((\kappa + \kappa_-)x) + a_+ \cosh((\kappa - \kappa_-)x) \right) , \tag{35}$$

$$\varphi_- = A_- \left( \cosh((\kappa_+ + \kappa)x) - a_- \cosh((\kappa_+ - \kappa)x) \right) , \tag{36}$$

$$\varphi = A \cdot \left( \sinh((\kappa_+ + \kappa_-)x) + a \sinh((\kappa_+ - \kappa_-)x) \right) . \tag{37}$$

The three-soliton solutions in the initial form have arbitrary values of their three parameters but we specify them so that they are exactly equal to the eigenvalues $\kappa_+$, $\kappa$, $\kappa_-$ of the operator $L$ from Eqs. (24) – (26). The coefficients in the above formulas are the following:

$$a_+ = \frac{\kappa + \kappa_-}{\kappa - \kappa_-}, \qquad a_- = \frac{\kappa_+ + \kappa}{\kappa_+ - \kappa}, \qquad a = \frac{\kappa_+ + \kappa_-}{\kappa_+ - \kappa_-} \tag{38}$$

$$A_+ = 2\kappa_+ \sqrt{a_- a}, \qquad A_- = 2\kappa_- \sqrt{a_+ a}, \qquad A = -2\kappa \sqrt{a_+ a_-} . \tag{39}$$

The function $F_0$ looks like

$$F_0 = \cosh((\kappa_+ + \kappa + \kappa_-)x) + a_+ a \cosh((\kappa_+ + \kappa - \kappa_-)x) + \\ + a_+ a_- \cosh((\kappa_+ + \kappa_- - \kappa)x) + a_- a \cosh((\kappa + \kappa_- - \kappa_+)x) \tag{40}$$

Using this function, we construct the potential well

$$U_0(x) = -2 \frac{d^2}{dx^2} \ln F_0 = -\frac{D_x^2 F_0 \cdot F_0}{F_0^2} \tag{41}$$

and the Schrödinger operator

$$L_0 = -\frac{d^2}{dx^2} + U_0(x). \tag{42}$$

Its normalized eigenfunctions naturally obey the equations

$$L_0 w_+ = -\kappa_+^2 w_+ , \quad L_0 w = -\kappa^2 w, \quad L_0 w_- = -\kappa_-^2 w_- . \tag{43}$$

and are built from the above functions (35) – (40) as follows

$$w_+ = \frac{1}{\sqrt{2\kappa_+}} \frac{\varphi_+}{F_0}, \quad w = \frac{1}{\sqrt{2\kappa}} \frac{\varphi}{F_0}, \quad w_- = \frac{1}{\sqrt{2\kappa_-}} \frac{\varphi_-}{F_0}. \tag{44}$$

These three functions, complemented by the continuous spectrum eigenfunctions $w_k(x)$, constitute the complete orthonormal basis. Note that the special choice of the coefficients in Eqs. (39) and (44) assures the function normalization to unity.

Dividing Eq. (34) by $F_0^2$ and using the definition (41) for $U_0(x)$, we obtain the explicit spatial dependence of the potential through the normalized eigenfunctions (44):

$$U_0(x) = -4(\kappa_+ w_+^2 + \kappa w^2 + \kappa_- w_-^2). \tag{45}$$

Note that the potential value at the origin and its integral are expressed simply through the eigenvalues:

$$U_0(0) = -2(\kappa_+^2 - \kappa^2 + \kappa_-^2) , \tag{46}$$

$$J_0 = \int_{-\infty}^{\infty} U_0(x) dx = -2(\kappa_+ + \kappa + \kappa_-). \tag{47}$$

Now we can conclude that two Hermitian operators $L$ and $L_0$ have the completely identical set of eigenvalues, since, besides the coincidence of three discrete levels, they have the same continuous spectrum $\varepsilon_k = k^2$ due to vanishing the potential at infinity. It means that there is the unitary operator $\mathbf{S}$ that transforms the orthonormal basis of eigenfunctions of operator $L_0$ to the orthonormal basis of operator $L$. In particular, the normalized eigenfunctions, corresponding to discrete levels, are connected by relations:

$$u_+(x) = \frac{1}{\sqrt{N_u}} u(x) = \mathbf{S} w_+, \quad \chi(x) = \mathbf{S} w(x), \quad v_-(x) = \frac{1}{\sqrt{N_v}} v(x) = \mathbf{S} w_-, \tag{48}$$

where the normalization constants are as follows

$$N_u = \int_{-\infty}^{\infty} u^2(x) dx , \quad N_v = \int_{-\infty}^{\infty} v^2(x) dx . \tag{49}$$

As known [30], after applying consequently the unitary operation to the operator $L_0$ we transform it to the operator $L = \mathbf{S} L_0 \mathbf{S}^{-1}$.

At last, using the eigenfunctions $u_+(x)$, $\chi(x)$ and $v_-(x)$ we can construct the potential $\widetilde{U}(x)$ as the analogue of formula (45) for the potential $U_0(x)$:

$$\widetilde{U}(x) = -4(\kappa_+ u_+^2 + \kappa \chi^2 + \kappa_- v_-^2). \tag{50}$$

In general these potentials are different, but, as it follows from the relations (48) between the eigenfunctions and properties of the unitary transformation, the integral $\widetilde{J}$ of the potential (50) is equal to $J_0$ of Eq. (47), and hence, we find for it the following useful expression:

$$\widetilde{J} = \int_{-\infty}^{\infty} \widetilde{U}(x) dx = -2(\kappa_+ + \kappa + \kappa_-). \tag{51}$$

In the next section, we compare all the potentials and find the explicit analytical approximation of the functions $u(x)$ and $v(x)$ for the above pairs of numerical values of the parameters of the strength of pumping and the dissipation coefficient.

### Quantitative description of static soliton bound states in the PDNLSE systems

The numerical solutions for static double-hump solitons and the effective procedure of finding the soliton complexes in the framework of Eqs. (6) have been found in Ref. 27. As shown in Ref. 28, a domain of existence of such solutions begins once the dissipation parameter is not zero and at the pumping strength values lying above the threshold curve $h_{th}(\gamma) > \gamma$. The stability region of these static solutions is located at large enough values of the dissipation, as it has been found by simulations of the PDNLSE, and it is partly confirmed by the direct stability analysis [25, 26, 28]. In this section, we intend to analyze the structure of the static solutions and to propose an analytical approximation for them without the variational approach. Therefore, we consider the solutions at the parameter values just below the stability boundary and inside the stability region. The choice of the parameters $h$ and $\gamma$ aims to compare our results with those of Refs. 26 and 27.

We have solved numerically Eqs. (22) and (23) and found the solutions in terms of the functions $u(x)$ and $v(x)$. These functions are presented below for values of parameters $h = 0.9$ and $\gamma = 0.565$ and $h = 1.05$ and $\gamma = 0.6$, respectively.

The first pair of the investigated parameters has been shown [27] to belong to the instability region, and it is placed just near the stability boundary of the double-hump soliton. After finding the solutions $u(x)$ and $v(x)$ for the parameter values $h = 0.9$ and $\gamma = 0.565$, using the transformation (17) and (18), we revert exactly to the results for the real and imaginary parts of the function $\psi(x)$ (see Fig.1 in Ref. 27). In terms of the functions $\psi_1$ and $\psi_2$ the static solutions look like the two-soliton bound state. On the other hand, considering the system of the Schrödinger–like equations (22) and (23) prompts the other kind of classification of two-component soliton solution. It is easy to see from Fig. 1, that $u(x)$ (black line 1) is the eigenfunction of the ground state with the eigenvalue $-\kappa_+^2$ and $v(x)$ (red line 2) is the eigenfunction with the eigenvalue $-\kappa_-^2$ of the second excited state of the Schrödinger operator $L$, respectively. Remembering the representation (16) and the properties of single solitons (3) and (4), we can classify the bound soliton structure as the superposition of complex generalizations of solitons $\psi_+$ and $\psi_-$, which differ in character of localization, in particular, by the fast and slow decaying asymptotics, respectively.

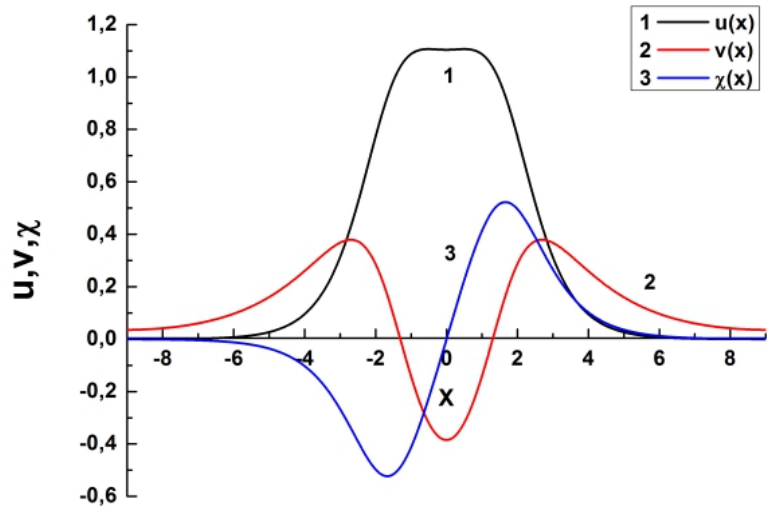

*Fig. 1. The numerical soliton solutions $u(x)$ ( black line 1) and $v(x)$ (red line 2) as the even eigenfunctions of the effective Schrödinger operator $L$. Its odd normalized function $\chi(x)$ denotes as the blue line 3. The parameters of the pumping strength and dissipation are equal as $h = 0.9$ and $\gamma = 0.565$.*

The numerical values of the functions $u$ and $v$ at the origin are the following: $u_0 = 1.10407$ and $v_0 = -0.38472$. We do not show more digits after the decimal point here and further but note that calculations performed have required a very high accuracy for the initial conditions and parameters. Using Eq. (24), we construct the potential well $U(x)$ for this case and show it as the black line 1 in Fig. 2. Now, we are able to find the eigenfunction $\chi(x)$ for the first excited state of the operator $L$ with the potential $U(x)$. The normalized odd eigenfunction $\chi(x)$ is shown in Fig. 1 as the blue line 3. Its eigenvalue $-\kappa^2$ is found numerically, and finally, we obtain the parameter $\kappa = 1.08625$. The calculated values of the parameters $\kappa_+$ and $\kappa_-$ of Eq. (5) are $1.30405$ and $0.54722$, respectively. With the knowledge of $\kappa_+$, $\kappa$ and $\kappa_-$ we can use explicit expressions (35) – (41) for the construction of the corresponding three-soliton solution and the potential well $U_0(x)$, and the normalized eigenfunctions $w_+$, $w$ and $w_-$. The potential $U_0(x)$, given by formula (41), is indicated by the red line 2 in Fig. 2. The proximity of the potentials $U(x)$ and $U_0(x)$, calculated numerically by Eq. (24) and analytically by Eq. (41), respectively, is quite well. This fact points out that the eigenfunctions (44) can be a very good first approximation for the eigenfunctions of the system (26) with the *right asymptotic behavior*. Now it is clear that the unitary operator $\mathbf{S} = \exp(i\mathbf{R})$, introduced in Eqs. (48), is close to the identity operator. Therefore the following operator expansions hold: $\mathbf{S} \cong \mathbf{I} + i\mathbf{R}$ and $\mathbf{S}^{-1} \cong \mathbf{I} - i\mathbf{R}$, and $L \cong L_0 - i[L_0, \mathbf{R}]$, where $\mathbf{R}$ is a "small" Hermitian operator [30] and the square bracket means the commutator. In result, it is natural to replace $u_+$ and $v_+$ by $w_+$ and $w_-$, respectively, and to represent for functions $u(x)$ and $v(x)$ in the form

$$u(x) = u_0 w_+(x)/w_+(0), \qquad v(x) = v_0 w_-(x)/w_-(0). \qquad (52)$$

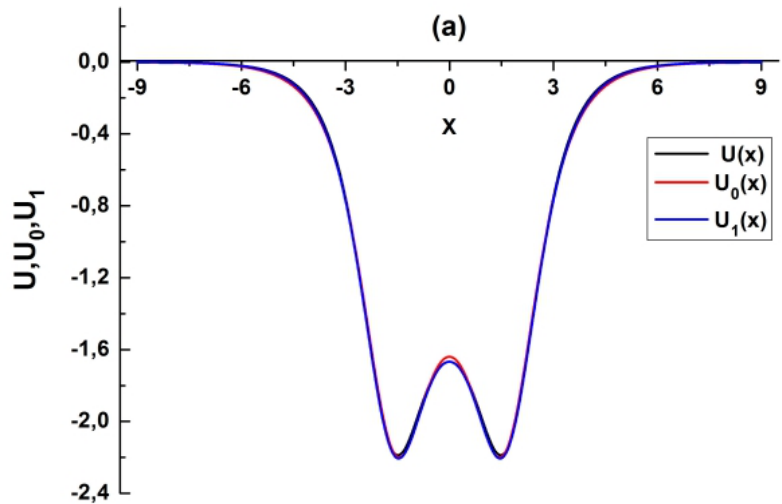

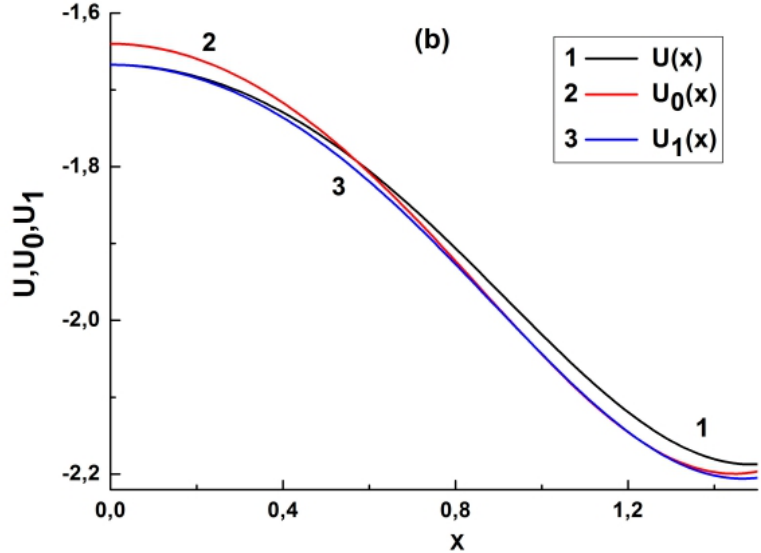

*Fig. 2. The general look (a) of the potential wells and their enlarge portions (b) between the origin and the minima for distinguishing the lines. The potential wells U (black line 1) and $U_0$ (red line 2) are for the Schrödinger operators L and $L_0$, respectively. The potential $U_1$ (blue line 3) is calculated in the one-mode approximation. The parameters of the pumping strength and dissipation are equal as $h = 0.9$ and $\gamma = 0.565$.*

We use the simple condition of the equality of amplitudes of the numerical solution and the analytical expression at the origin. Then we put them into Eq. (24) and find the approximate form of the potential well $U_1(x)$. This potential is shown as the blue line 3 in Fig. 2, and it reproduces quiet well the numerical result and is distinctly better than the curve $U_0(x)$ at the origin and at $|x| > 3$.

By the next step, we calculate the constants $N_u$ and $N_v$ and find the normalized eigenfunctions $u_+(x)$ and $v_-(x)$ from Eq. (44). With the addition of the eigenfunction $\chi(x)$ we construct the potential $\widetilde{U}(x)$ by the use of the formula (50) and make sure that it differs from $U_0(x)$ to the same extent as the orthonormal basis functions of operators $L$ and $L_0$, being related by the unitary transformation, differ each other. However, it turns out remarkably that the potential $\widetilde{U}(x)$ practically coincides with $U(x)$ obtained from Eq. (24) and shown in Fig. 2 by the black line. Consequently, the integral $J$ of the potential $U(x)$ can be estimated as $J = \widetilde{J} = J_0$.

The second pair of the parameters $h = 1.05$ and $\gamma = 0.6$ is placed inside the stability region, and numerical simulation demonstrates surviving the static double-hump solution in the PDNLSE [26]. We repeat the above procedure of finding the potentials $U$, $U_0$ and $U_1$ for this couple of the strength of pumping and dissipation coefficient. The results of the numerical solution of the system of Eqs. (22) and (23) for the functions $u(x)$ and $v(x)$ are presented in Fig. 3. Now, the function $u(x)$ (black line 1) has the double-hump shape and decays much faster than the function $v(x)$ (red line 2) because of a large difference of $\kappa_+ = 1.36443$ and $\kappa_- = 0.37191$. The numerical values for the functions $u$ and $v$ at the origin are the following: $u_0 = 0.77424$ and $v_0 = -0.23201$. We construct the potential well $U(x)$ for this case by using Eq. (24) and present it as the black line in Fig. 4. Then, we find the odd eigenfunction $\chi(x)$ for the first excited state of the operator $L$ with the obtained potential $U(x)$. The normalized eigenfunction $\chi(x)$ is shown in Fig. 3 as the blue line 3. Its eigenvalue $-\kappa^2$ is found numerically and, due to the deep double-well form of the potential $U(x)$ the value of $\kappa = 1.26341$ appears to be close to the parameter $\kappa_+$. As before, with the knowledge of $\kappa_+$, $\kappa$ and $\kappa_-$ we employ Eqs. (35) – (41) for construction of the three-soliton solution and the potential well $U_0(x)$, as well as the normalized eigenfunctions $w_+$, $w$ and $w_-$. The potential $U_0(x)$, found from formula (41), is shown as the red line in Fig. 4. The potentials $U(x)$ and $U_0(x)$ are close enough but definitely different at the origin. Therefore, the analytical formulas (35) – (41) do not give the exact solution of the system (26) about what one could think for the pair of parameters $h$ and $\gamma$ close to the previous first due to the fine proximity of the corresponding potentials. Nevertheless, the eigenfunctions (44) appear to be again a good approximation for the eigenfunctions of the system (26) owing to the right asymptotic behavior. We use the expressions (52) for functions $u(x)$ and $v(x)$ and Eq. (24) in order to find the approximate potential well $U_1(x)$. The latter is shown as the blue line in Fig. 4 and reproduces quiet well the numerical result $U(x)$.

After finding the normalized eigenfunctions $u_+(x)$ and $v_-(x)$ and $\chi$ from Eq. (48) for the given pair of the parameters $h$ and $\gamma$ we construct the potential $\tilde{U}(x)$ using the formula (50) and make sure that it does not differ in the eye from the potential $U(x)$ presented in Fig. 4 by the black line.

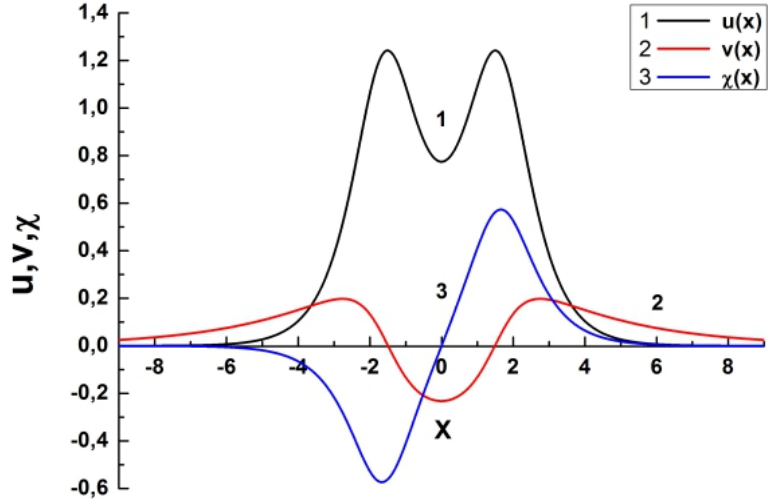

*Fig. 3. The numerical soliton solutions $u(x)$ ( black line 1) and $v(x)$ (red line 2) as the even eigenfunctions of the effective Schrödinger operator $L$. Its odd normalized function $\chi(x)$ denotes as the blue line 3. The parameters of the pumping strength and dissipation are equal as $h = 1.05$ and $\gamma = 0.6$.*

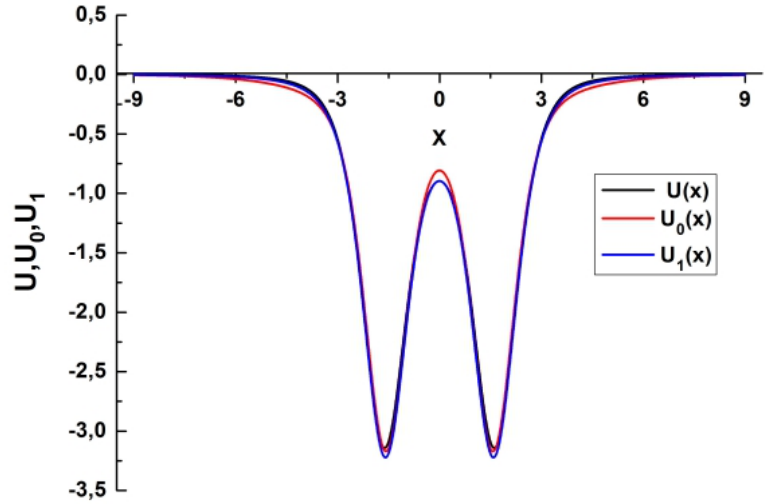

*Fig. 4. The potential wells U (black line) and $U_0$ (red line) are for the Schrödinger operators L and $L_0$, respectively. The potential $U_1$ (blue line) is calculated in the one-mode approximation. The mutual arrangement of the lines are the same as in Fig. 2. The parameters of the pumping strength and dissipation are equal as $h = 1.05$ and $\gamma = 0.6$.*

The dependences $U_1(x)$ in Fig. 2 and Fig. 4 are results of the one-mode approximation of solutions $u(x)$ and $v(x)$. Indeed, the eigenfunctions of the operator $L$ can be expanded by using the orthonormal basis functions of the operator $L_0$ as follows

$$u_+(x) = B_0 w_+ + B_2 w_- + \int_{-\infty}^{\infty} B_k w_k(x) dk \tag{53}$$

$$v_-(x) = C_0 w_- + C_2 w_+ + \int_{-\infty}^{\infty} C_k w_k(x) dk \tag{54}$$

The terms $B_1 w$ and $C_1 w$ are absent in the expansions (51) and (52), respectively, since we consider the even functions $u(x)$ and $v(x)$. We have used so far only the first terms in these expansions. The next step would be taking into account also terms $B_2 w_-$ and $C_2 w_-$. This two-mode approximation with using only two localized eigenfunctions $w_+$ and $w_-$ needs notice. It concerns the problem of the asymptotic behavior of the functions. All is normal in the case of the function $v(x)$, whose slowly decaying asymptotics is the same as that of the eigenfunction $w_-$. However, it is not the case for the function $u(x)$. This simply means that at large $x$ the contribution of the eigenfunction $w_-$ in the expansion (53) must be compensated by the integral term, i.e., by the contribution of continuous spectrum states. As an example of a similar situation, it is enough to present the expansion of the normalized ground state eigenfunction $\varphi_0 = \sqrt{3/4}\,\text{sech}^2 x$ of the Schrödinger operator with the potential $-6\,\text{sech}^2 x$ by the use of the orthonormal basis of the operator with the potential $-2\,\text{sech}^2 x$, having the ground state eigenfunction $\psi_0 = \sqrt{1/2}\,\text{sech}\,x$ (cf. Eqs. (7) and (8)):

$$\varphi_0 = \frac{\sqrt{3}}{8}\pi \cdot \psi_0 + \left(\varphi_0 - \frac{\sqrt{3}}{8}\pi \cdot \psi_0\right). \tag{55}$$

The expression in brackets is the contribution of the continuous spectrum waves, and it is fully localized and compensates completely for the contribution of the ground state function $\psi_0$. Another situation, when the eigenfunction is expanded only into a few localized states, is rather rare. As an example, we mention the expansion of the function $\psi_0$ into only two even localized eigenfunctions of the Schrödinger operator with the potential $-12\,\text{sech}^2 x$. We refer to these simple findings because, unfortunately, the explicit form of the continuous spectrum eigenfunctions of the operator (41) for the case of the three-soliton solutions are not available yet, as to our knowledge, except asymptotics and other scattering data.

Keeping in mind the above notice, we restrict ourselves by the two-mode approximation with truncated expansions (53) and (54) without the contribution of the continuous spectrum states. For both pairs of the parameters $h$ and $\gamma$ the coefficients $B_0$, $B_2$, $C_0$ and $C_2$ are calculated, as usual, by multiplying the Eqs. (53) and (54) by the corresponding analytical eigenfunction and by numerical finding the scalar products. For values $h = 0.9$ and $\gamma = 0.565$ we find $B_0 = 0.99999$ and $B_2 = 0.00022$, and $C_0 = 0.99928$ and $C_2 = -0.00023$. Thus, indeed the unitary operator $\mathbf{S}$ in Eqs. (47) is very close to the identity operator. The function $u_+(x)$ of Eq. (48), obtained from $u(x)$ (black line 1 in Fig.1), is entirely reproduced by the analytical expression Eq. (53) with the found coefficients. Whereas the function $v_-(x)$ of Eq. (48) as the normalized curve $v(x)$ (red line 2 in Fig.1) is indistinguishable to the eye from the analytical expression of Eq. (54) with the found coefficients up to regions with $x \cong \pm 7$. At a larger distance $u(x)$ vanishes and asymptotics of $v(x)$ tends to those of the single soliton $\psi_-(x)$ of Eq. (4). The proximity of the ratios $\alpha_1 = B_2/B_0$ and $\alpha_2 = -C_2/C_0$ is not accidental. It is easy to see that the orthogonality of $u(x)$ and $v(x)$ in the two-mode approximation leads to the exact equality $\alpha_1 = \alpha_2$. At last, we adduce the corresponding results for the pair of parameters $h = 1.05$ and $\gamma = 0.6$. The calculated coefficients, in this case, are as following: $B_0 = 0.99995$ and $B_2 = 0.00365$, and $C_0 = 0.99318$ and $C_2 = -0.00380$. As a result, the function $u_+(x)$, obtained by the normalization of $u(x)$ (black line 1 in Fig.3), is described very well by analytical approximation, whereas the function $v_-(x)$ differs from analytical expression by a few percents at the origin and less at its maxima. Hence it occurs that the two-mode approximation is closer to the numerical solution $v(x)$, than the one-mode approximation of Eq. (50), at intermediate distances.

## Conclusion

The main findings of this study are as follows:

1. We have described quantitatively the static double-hump solution of the parametrically driven damped nonlinear Schrödinger equation for the soliton bound state by the explicit three-soliton formula of the integrable equations. This is the unexpected and intriguing result that nonlinear system with large enough dissipation can be described quite well by the exact multisoliton formula of the integrable equations.

2. As a consequence, the static soliton structure, which is usually interpreted as two-soliton bound state, is constructed as the "nonlinear superposition" of three solitons with three fixed parameters, which depend on the strength of the parametric pumping and the dissipation coefficient and determine all the structure configuration. It suggests a new regard on the internal structure of the double-hump solitons and proposes a new set of internal degrees of freedom of the soliton complex in the parametrically driven and damped systems. This makes it possible to formulate further a new variational ansatz for the description the complex internal dynamics, including regular and chaotic oscillation regimes of the constituting solitons.

3. The performed transformation of the governing system of equations to the Schrödinger-like equations reveals the close relation between the structural parts of the double-hump soliton configuration with different decaying laws and the eigenfunctions of the Schrödinger operator with the symmetric potential well. This allows us to consider the even eigenfunctions of effective spectral problem as the principal soliton modes. It gives rise to a new way of classification of static multisoliton complexes found numerically before. It is clear that they are constructed from pair of eigenfunctions of the Schrödinger-like operator, corresponding to its ground and higher excited states.

4. The explicit form of the unitary operator, which transforms the orthonormal basis functions in the present study, is still an open problem. There is the well-known example of the transformation of the reflectionless potential well of the Schrödinger operator in the direct scattering problem associated with the Korteweg–de Vries equation, which does not change the eigenvalue set. However, in this case, the potential loses its symmetric form under the transformation in contrast to what we encounter in the present study.

The new tasks emerging from awareness of the above intriguing results will be subjects of forthcoming works.

ACKNOWLEDGMENTS

This work was performed using computational facilities of grid-cluster ILTPE – B.Verkin Institute for Low Temperature Physics and Engineering of the National Academy of Sciences of Ukraine. This research was funded in part by the Luxembourg National Research Fund (FNR) grant C19/MS/13718694/QML-FLEX. For the purpose of open access, O.C. has applied a Creative Commons Attribution 4.0 International (CC BY 4.0) license to any Author Accepted Manuscript version arising from this submission.References

1. A.S. Davydov and N.I. Kislukha, *Phys. Stat. Sol. B* **59**, 465 (1973).
2. A.S. Davydov, *Physica Scripta* **20**, 387 (1979).
3. L.S. Brizhik and A.S. Davydov, *Fiz. Nizk. Temp.* **10,** 748 (1984).
4. L.S. Brizhik and A.A. Eremko, *Physica D* **81**, 295 (1995).19